\documentclass[runningheads]{llncs}

\usepackage{fontspec}
\usepackage[frozencache,cachedir=minted-cache]{minted}
\usepackage{graphicx}
\usepackage{multicol}
\usepackage{amsmath}
\usepackage{amssymb}
\usepackage{mathtools}
\usepackage{bcprules}
\usepackage{hyperref}
\usepackage{booktabs}
\usepackage{xcolor}
\usepackage{tcolorbox}
\usepackage{tikz}
\usetikzlibrary{positioning}

\usepackage{color}

\definecolor{codebg}{RGB}{250,250,250}
\definecolor{codeframe}{RGB}{220,220,220}
\setminted{
  fontsize=\footnotesize,
  bgcolor=codebg,
  frame=leftline,
  framesep=2mm,
  framerule=0.4pt,
  rulecolor=codeframe,
  numbersep=8pt,
  linenos,
  breaklines
}

\newmintinline[lean]{lean4}{bgcolor=white}
\newminted[leancode]{lean4}{}

\newcommand{\mknote}[3]{%
  \tcbox[
    on line,
    arc=3pt,
    colback=#1,
    colframe=#1,
    fontupper=\bfseries\footnotesize\ttfamily,
    boxsep=1pt,
    left=2pt,
    right=2pt,
    top=0.5pt,
    bottom=0.5pt
  ]{\color{white}#2}: {\color{#1}\textbf{#3}}%
}
\renewcommand{\mknote}[3]{}

\newcommand{\smallwedge}{\mathrel{\!\text{\raisebox{0.7ex}{\scalebox{0.6}{$\wedge$}}}\!}}
\newcommand\capt{\smallwedge}

\newcounter{RuleRef}
\newcommand{\ruledef}[1]{%
  \refstepcounter{RuleRef}\textsc{#1}\label{rule:#1}}
\newcommand{\ruleref}[1]{\textsc{(\hyperref[rule:#1]{#1})}}

\newcounter{RRuleRef}
\newcommand{\rruledef}[1]{%
  \refstepcounter{RRuleRef}\textsc{#1}\label{rrule:#1}}
\newcommand{\rruleref}[1]{\textsc{(\hyperref[rrule:#1]{#1})}}

\makeatletter%
\begin{document}
\title{Agentic Proof Automation: A Case Study}
\author{Yichen Xu\orcidID{0000-0003-2089-6767}\and
Martin Odersky\orcidID{0009-0005-3923-8993}}%
\authorrunning{Y. Xu et al.}
\institute{EPFL, Lausanne, Switzerland\\ %
\email{yichen.xu@epfl.ch} \quad \email{martin.odersky@epfl.ch}}
\maketitle              %
\begin{abstract}
Proof engineering is notoriously labor-intensive:
proofs that are straightforward on paper often require lengthy scripts in theorem provers.
Recent advances in large language models (LLMs) create new opportunities for proof automation:
modern LLMs not only generate proof scripts,
but also support agentic behavior,
exploring codebases and iteratively refining their outputs against prover feedback.
These advances enable an emerging scheme
where LLM-based agents undertake most proof engineering under human guidance.
Humans provide mathematical insight (definitions, theorems, proof strategies);
agents handle the mechanical work of proof development.
We call this scheme \emph{agentic proof automation}.

We present this scheme through a case study:
mechanizing the semantic type soundness of 
a sophisticated formal system, System Capless, in Lean 4,
comprising over 14,000 lines of code.
Using off-the-shelf LLM agents with a single lightweight proof-checking tool,
the agents completed 189 proof engineering tasks with an 87\% success rate,
only 16\% requiring human intervention.
The case study demonstrates that agents are capable proof engineers
that substantially boost productivity,
though they fall short in creative reasoning and still require human guidance in certain cases.
We release an interactive explorer where readers can examine all agent interactions;
the mechanization is open-sourced for experiments and extensions.

\end{abstract}

\section{Introduction}

Machine-checked proofs, or mechanized proofs,
provide highly-reliable guarantees for security-critical software.
Systems verified through interactive theorem provers (ITPs)
such as Coq \cite{Coq2020}, Lean \cite{lean2017}, or Isabelle/HOL \cite{isabelle-hol-2002}
offer correctness properties checked by machines.
These guarantees are more reliable than testing, 
which can miss edge cases,
and traditional pencil-and-paper proofs, which are prone to human error.

However, proof mechanization has been notoriously difficult.
A substantial gap exists between
the informal mathematical arguments that humans sketch on paper or have in mind
and the machine-checked proofs that theorem provers demand.
For instance, CompCert~\cite{leroy2008compcert}, a verified C compiler, took six person-years to mechanize in approximately 100,000 lines of Coq.
When a human writes ``this follows by straightforward induction'',
this \emph{straightforward}ness can often require hundreds of lines of justification in proof assistants.
This translation demands not only deep expertise in both the problem domain and interactive theorem provers,
but also considerable time investment.
For a complex system,
proof development can consume months or years of expert effort.
This makes formal verification prohibitively expensive for most real-world software.

Proof automation is proposed to narrow this gap,
and researchers have developed various techniques~\cite{Blaauwbroek2020Tactician,Blanchette2013SledgehammerSMT,Boehme2010Sledgehammer,Limperg2023Aesop,Gauthier2021TacticToe}.
Tactics like \texttt{eauto} and \texttt{try} discharge routine proof obligations automatically.
Highly automated tools like Stainless~\cite{Hamza2019Stainless} and Liquid Haskell~\cite{Vazou2014LiquidHaskell}
integrate SMT solvers to handle verification tasks.
These advances improve productivity, but substantial manual effort remains necessary.
Even with sophisticated automation,
proof engineers sometimes still need to write lengthy scripts
that manually manage proof states.
Moreover, existing automation offers limited recoverability or interpretability:
when a hammer or tactic fails,
users receive little insight into why,
and cannot easily guide the tool toward a successful proof.
The burden of proof engineering persists.

Recent advances in large language models (LLMs)~\cite{vaswani2017attention,devlin2018bert,brown2020language,radford2019language,touvron2023llama,chowdhery2022palm} offer new opportunities to address this challenge.
Modern LLMs demonstrate capabilities in generating code, which includes proof scripts.
These models have been trained on vast corpora that 
include both natural and formal languages.
More importantly, these models now support agentic behaviors:
they can explore codebases autonomously,
generate and test code,
backtrack from failed attempts,
and iteratively refine their outputs based on feedback.
For proof development, this means LLMs can interact with theorem provers in a feedback loop:
they inspect proof goals,
generate proof scripts,
receive possible error messages from the prover,
and refine their attempts accordingly.
This interactive capability transforms LLMs from mere code generators into active proof engineers.
Unlike traditional proof automation that operates purely at the formal level,
LLMs bridge natural and formal languages,
potentially serving as translators between human mathematical intuition and machine-checked proofs.

Our experience with LLM-based agents shows their impressive capabilities in proof engineering.
Agents can search through related modules to find relevant lemmas,
generate proof scripts to close holes in proofs,
recursively refine their attempts until they succeed.
Otherwise, they point out flawed assumptions,
or suggest lemma decomposition when the goal is too complicated to be directly proven.
These capabilities enable a new proof development scheme
in which LLM agents undertake most proof engineering work under human guidance.
In this scheme,
humans focus on the key mathematical insights:
they write theorem statements, lemma decompositions,
and natural language descriptions of proof strategies.
The LLM agent then generates the proof scripts to fill the holes in the proof,
managing most of the technical details of the theorem prover,
and pointing out flaws in the human's reasoning when it finds any.
We call this emerging scheme \emph{agentic proof automation}.

This paper aims to demonstrate this scheme with a case study:
the mechanization of the soundness
of a sophisticated formal system (System Capless \cite{Xu2025WhatsInTheBox})
in Lean 4,
comprising over 14,000 lines of code.
Humans defined the formal system and wrote the definitions,
while agents handled proof development based on high-level guidance.
We collected an extensive record of the interactions between the human and the agent
and manually annotated them into \emph{tasks},
which represent a minimal unit of human-agent interaction
where the human expresses an intention and the agent attempts to fulfill it.
Each task is categorized by its nature (e.g., finishing a proof, repairing a broken proof, refactoring code)
and evaluated based on its outcome (success, partial success, failure).
We report concrete examples of human-agent interactions,
quantitative analysis of the agents' contributions including success rates and performance patterns,
and insights on effectively leveraging this new scheme for proof development.
This case study is accompanied by an explorer
that allows readers to interactively explore 
the raw interactions and annotated tasks,
seeing how the human provided guidance
and how the agent engineered the proofs in a feedback loop.
This explorer can either be deployed locally from the open source repository\footnote{\url{https://github.com/linyxus/apa-explorer}}
or accessed online\footnote{\url{https://apa.univalence.xyz/}}.
We refer to examples on this explorer throughout the paper.
The case study mechanization is also open-sourced\footnote{\url{https://github.com/Linyxus/semantic/tree/case-study}}.

Agentic proof automation lowers the barrier to machine-verified proof development.
Humans can explore formal properties of their systems through natural language guidance,
leaving low-level proof details to agents.
This shift could expand formal methods beyond niche applications:
as verification becomes accessible to practitioners without years of theorem prover expertise,
mathematically rigorous software development becomes practical for domains that currently rely on testing alone.
These potential impacts motivate this case study,
which explores agentic proof automation in practice and demonstrates its viability for real-world proof development.
 
\section{Agentic Proof Automation By Examples}

This section illustrates \emph{agentic proof automation} in action.
We describe the setup, present concrete examples from our case study,
analyze the patterns that emerged, and discuss the workflow's characteristics.

\subsection{Setup}

Our case study uses off-the-shelf LLM agents
equipped with standard software engineering tools.
We did not fine-tune models or build custom proof-generation systems.
Instead, we used general-purpose coding agents
and added one lightweight tool for Lean 4 integration.
This simple setup has two implications.
First, any practitioner can adopt the workflow without specialized infrastructure.
The mechanization is open-sourced, so that others can
experiment with the workflow or use it as a starting point for their own projects.
Second, the workflow generalizes:
it transfers to other proof assistants
and scales with improvements in frontier models.

\paragraph{Models and Frameworks.}
We used several frontier LLMs throughout the project:
Claude Sonnet 4.5, Claude Opus 4.5, Claude Opus 4.1, and GPT Codex 5.1.
These models ran within two agentic frameworks:
Claude Code (for Claude models) and Codex-CLI (for GPT models).
Both frameworks provide similar capabilities:
they wrap the underlying LLM with tools for file system access,
command execution,
and iterative refinement based on tool outputs.

\paragraph{Agent Tools.}
The agents have access to four categories of tools:
\begin{itemize}
\item \emph{Code exploration}: searching for files by pattern,
  grepping for definitions and usages,
  and reading file contents.
\item \emph{File modification}: creating and editing files,
  with the ability to make targeted changes to specific locations.
\item \emph{Lean 4 building}: a custom tool called \texttt{lean4check}
  that compiles a Lean 4 module and renders the result.
  This tool reports success, or formats error messages
  with enclosing source contexts for easier interpretation.
\end{itemize}
The \texttt{lean4check} tool is the only Lean-specific addition.
It invokes the Lean compiler on the command line,
collects its output,
and renders errors alongside their source context
so the agent can interpret them more easily.
The tool requires no integration with the proof assistant's internals.
For example, when a proof script has a type mismatch error, \texttt{lean4check} produces:
\begin{verbatim}
ERROR: Semantic/CC/Eval/SmallStep.lean:35:45
33 |   (hv : Exp.IsVal v) ->
34 |   h l = none ->
35 |   Step h (.letin v e) (h.extend l v) e
   |                                      ^
   | Application type mismatch: argument
   |   e
   | has type
   |   Exp ({},x)
   | but is expected to have type
   |   Exp {}
\end{verbatim}
The output shows both the error message and the surrounding source code,
which provides better context for the agent to understand the issue.

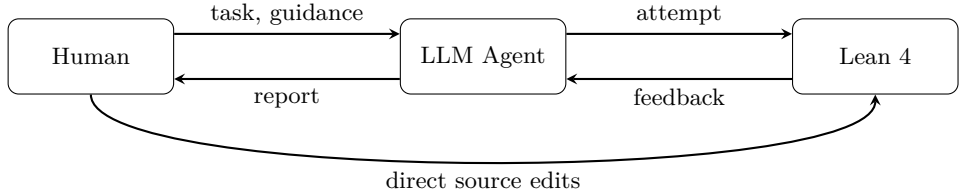
\begin{figure}[t]
\centering
\begin{tikzpicture}[
    node distance=2.5cm and 3cm,
    box/.style={rectangle, draw, rounded corners, minimum width=2.2cm, minimum height=1cm, align=center, font=\small},
    arrow/.style={->, >=stealth, thick},
    label/.style={font=\footnotesize, align=center}
]
    \node[box] (human) {Human};
    \node[box, right=of human] (agent) {LLM Agent};
    \node[box, right=of agent] (lean) {Lean 4};

    \draw[arrow, transform canvas={yshift=0.3cm}] (human.east) -- (agent.west)
        node[midway, above, label] {task, guidance};

    \draw[arrow, transform canvas={yshift=-0.3cm}] (agent.west) -- (human.east)
        node[midway, below, label] {report};

    \draw[arrow, transform canvas={yshift=0.3cm}] (agent.east) -- (lean.west)
        node[midway, above, label] {attempt};

    \draw[arrow, transform canvas={yshift=-0.3cm}] (lean.west) -- (agent.east)
        node[midway, below, label] {feedback};

    \draw[arrow] (human.south) .. controls +(0,-1.2) and +(0,-1.2) .. (lean.south)
        node[midway, below, label] {direct source edits};

\end{tikzpicture}
\caption{The human-agent-prover workflow. The human provides tasks and guidance. The agent generates proof scripts and refines them based on Lean's feedback, reporting success or identified obstacles back to the human.}
\label{fig:workflow}
\end{figure}

\paragraph{Workflow.}
A typical interaction proceeds as follows.
The human writes definitions, states theorems,
and optionally provides informal proof strategies as comments.
The agent then explores the codebase to build context:
it searches for related definitions, reads relevant files,
and identifies lemmas that might be useful.
When ready, the agent generates a proof script by modifying the file,
and invokes \texttt{lean4check}.
If the proof fails, the agent reads the error message,
interprets it, and refines its attempt.
This loop continues until the proof succeeds
or the agent determines that the goal requires human intervention: for instance,
when a necessary lemma is missing or the proof strategy is flawed.
Figure~\ref{fig:workflow} 
illustrates this interaction.

\subsection{A Brief Introduction to System Capless}

Our case study centers on mechanizing the type soundness of System Capless,
a calculus for capture-based effect tracking.
We briefly introduce this system here.

System Capless~\cite{Xu2025WhatsInTheBox}
is a foundational calculus for \emph{capture tracking}~\cite{boruchgruszecki_capturing_2023},
a type-systematic approach that tracks effects and resources
by recording which \emph{capabilities} a value may capture.

The key notion in capture tracking is \emph{capturing types},
which augment ordinary types with \emph{capture sets}.
A capturing type $S\capt C$ combines a shape type $S$
with a capture set $C$ that over-approximates the capabilities the value may reference.
For example, a function that performs console I/O
might have type $(\mathsf{String} \to \mathsf{Unit})\capt\{\mathsf{console}\}$,
indicating it captures the console capability, or in other words,
may perform console operations when invoked.

The mechanization establishes the \emph{semantic type soundness}~\cite{timany2024logical-type-soundness} of System Capless
using a logical relations approach.
Types are interpreted as predicates over memory states and expressions:
a value satisfies a type if it behaves according to the type's specification
under all possible heap extensions.
The central result is the \emph{fundamental theorem}
which shows that syntactic typing derivations imply
semantic typing.
This together with the adequacy of the type denotations
establishes that
\begin{itemize}
\item well-typed programs do not get stuck during evaluation, which is the classical type safety property, and
\item the capabilities accessed during execution are upper-bounded by the capture sets predicted by the type system.
\end{itemize}

The complexity of the mechanization stems from two sources.
First, System Capless itself is an expressive and sophisticated calculus
(full definitions in Appendix~\ref{sec:appendix-capless}).
It features universal and existential quantification over capture sets.
It also has \emph{dependent function types},
where the result type of a function may mention its parameter.
For example, a method that creates a logging function from a file capability
has type $\forall(f\!: \mathsf{File}). (\mathsf{String} \to \mathsf{Unit})\capt\{f\}$:
the returned logger captures exactly the file $f$ that was passed in.
This dependency is essential for capture tracking,
since the capabilities captured by a function's result
often depend on what was passed as an argument.
Second, the semantic type soundness approach
demands in-depth reasoning about
the type system,
the runtime semantics,
and the type denotations (which are logical relations that interpret types as predicates).
This interconnected reasoning makes the proofs intricate.

The mechanization consists of over 14,000 lines of Lean 4 code,
organized around de Bruijn indexed syntax with three kinds of variables
(term, type, and capture variables),
a big-step operational semantics with an explicit heap model,
and denotational semantics that interpret types as logical relations
parameterized by capability sets and memory states.

\subsection{Examples}
\label{sec:examples}

We now present distinct kinds of examples that arose during this case study,
each illustrating different aspects of human-agent collaboration.

\subsubsection{Proof Script Development}

The most common interaction pattern is proof script development:
the human states a theorem and the agent produces the proof.
We illustrate with an example from the System Capless mechanization.
The mechanization defines two operational semantics.
\texttt{Eval C m e Q} is big-step evaluation:
expression \texttt{e} in memory \texttt{m} eventually produces a result
satisfying postcondition \texttt{Q}, using capabilities \texttt{C}.
\texttt{Step C m1 e1 m2 e2} is small-step reduction:
expression \texttt{e1} takes one atomic step to \texttt{e2},
possibly updating memory from \texttt{m1} to \texttt{m2}.
Readers need not understand the full details of these definitions to follow the example.
Interested readers may refer to the attached artifact for fully documented definitions.

The theorem \texttt{step\_preserves\_eval} connects these two semantics:
if \texttt{e1} evaluates to \texttt{Q}, and \texttt{e1} steps to \texttt{e2},
then \texttt{e2} also evaluates to \texttt{Q}.
The human stated this theorem with an empty proof:
\begin{leancode}
theorem step_preserves_eval
    (heval : Eval C m1 e1 Q)
    (hstep : Step C m1 e1 m2 e2) :
    Eval C m2 e2 Q := by sorry
\end{leancode}
The prompt was simply: ``prove theorem \texttt{step\_preserves\_eval}.''

The agent began by reading the file and examining related definitions.
It then identified the proof structure:
``Let me start proving it by induction on the evaluation derivation.''
The first edit replaced the \texttt{sorry} with a proof skeleton:
\begin{leancode}
theorem step_preserves_eval ... := by
  induction heval generalizing m2 e2 with
  | eval_val hv hQ =>
    -- Values cannot step: contradiction
    exact absurd hstep (step_val_absurd hv)
  | eval_var hQ =>
    exact absurd hstep step_var_absurd
  | eval_apply => sorry
  | eval_invoke => sorry
  ...
\end{leancode}
The agent correctly identified that base cases
(\texttt{eval\_val}, \texttt{eval\_var}) lead to contradictions:
values and variables are answers that cannot step further.
It discovered the lemmas \texttt{step\_val\_absurd} and \texttt{step\_var\_absurd},
which state that values and variables cannot possibly take a reduction step,
by exploring the codebase.

What followed was an edit-check-refine loop,
including 23 edits to the proof script and 18 invocations of the Lean compiler.
The first 17 checks returned errors,
of which each guided the next refinement.
Consider the \texttt{eval\_letin} case, which required nested case analysis.
When a let-expression \texttt{let x = e1 in e2} steps,
three situations arise:
the bound expression \texttt{e1} steps (apply induction hypothesis),
\texttt{e1} is a variable (rename and continue),
or \texttt{e1} is a value (allocate the value in memory).
The agent handled each:
\begin{leancode}
| eval_letin hpred heval_e1 h_nonstuck h_val h_var ih =>
  cases hstep with
  | step_ctx_letin hstep_e1 =>
    have heval' := ih hstep_e1
    have hsub := step_memory_monotonic hstep_e1
    apply Eval.eval_letin hpred heval' h_nonstuck ...
  | step_rename => ...
  | step_lift => ...
\end{leancode}
On the 18th compiler invocation, the proof compiled, totaling 240 lines.
The entire interaction, from prompt to complete proof,
took one exchange with no human intervention.
The agent discovered relevant lemmas (\texttt{step\_memory\_monotonic},
\texttt{eval\_ans\_holds\_post}),
identified the induction principle,
and handled all cases with appropriate tactics.

\subsubsection{Proof Repair After Refactoring}\label{sec:proof-repair}

When definitions change, downstream proofs often break.
Agents can systematically repair these proofs,
and sometimes identify deeper issues in the process.
In this example,
the human extended the calculus with mutable references,
adding \texttt{read} and \texttt{write} operations to the syntax and semantics.
This broke proofs throughout the codebase.
The human asked the agent to repair \texttt{Semantics/Props.lean}:
\begin{quote}
``Fix all errors in Props.lean caused by this mutable reference refactor.''
\end{quote}

The agent worked through the file systematically,
making 27 edits and invoking the Lean compiler 18 times
over 14 minutes.
It fixed ten distinct issues,
primarily adding cases for the new operations
and updating proofs to match changed definitions.

However, the agent encountered cases it could not resolve.
One theorem states that reduction is preserved
when we hide unused capabilities:
\begin{leancode*}{linenos=false}
theorem step_masked (hstep : Step C m1 e1 m2 e2) :
    Step C (m1.masked_caps C) e1 (m2.masked_caps C) e2
\end{leancode*}
Informally: if a step uses only capabilities in \texttt{C},
it should still work after hiding everything outside \texttt{C}.
When adding cases for \texttt{read} and \texttt{write},
the agent discovered the proof was impossible:
\begin{quote}
``For \texttt{step\_masked} to work with \texttt{step\_read/write},
the mutable cell must be in the capability set \texttt{C}.
But the current step rules don't require this!''
\end{quote}
The original semantics allowed reading any mutable cell,
but masking would hide cells outside \texttt{C}.
The agent marked these cases with \texttt{sorry} and reported the issue.
The human confirmed the diagnosis
and directed the agent to fix the semantics,
adding the missing membership requirement to \texttt{read} and \texttt{write}.

This example shows that proof repair is not purely mechanical.
The agent discovered that certain proofs were not just broken
but \emph{unprovable} under the current definitions,
and it identified the conceptual fix required.

\subsubsection{Answering Questions About the Mechanization}

Agents can also help navigate and understand the codebase.
Here is a concrete example.
The human posed a question about the mechanization:
\begin{quote}
``Does there exist a theorem which states that
as long as \texttt{Eval C m1 e1 Q},
there exists \texttt{m2} and \texttt{e2},
where \texttt{e2} is an answer, such that \texttt{Q m2 e2}?''
\end{quote}
The human wanted to know whether an adequacy theorem
which states that evaluation always terminates with an answer satisfying the postcondition
had been mechanized.

The agent searched the codebase with 10 tool calls
(6 greps, 3 file reads, 1 glob)
over 52 seconds, then reported:
\begin{quote}
``Based on my search through the CC module,
\textbf{no such theorem currently exists} in the codebase.
The closest related theorems I found are:
\texttt{eval\_ans\_holds\_post}, which says if \texttt{e} is already an answer
and has an evaluation, then the postcondition holds for it.
However, it requires \texttt{e} to already be an answer.
[\ldots]
The theorem you're asking about would be:
\begin{leancode*}{linenos=false}
theorem eval_exists_answer (heval : Eval C m1 e1 Q) :
    ∃ m2 e2, e2.IsAns ∧ Q e2 m2
\end{leancode*}
This would be an adequacy/termination theorem.
Such a theorem does not currently exist in the module.''
\end{quote}
The agent answered the question,
identified related theorems,
and formulated the statement the human had in mind.

\subsubsection{Stating and Proving Standalone Properties}

Continuing from the previous example,
the human asked the agent to state and prove the missing theorem:
\begin{quote}
``State and prove this theorem in Semantics/Props.lean.''
\end{quote}
The agent formalized the statement (refining the version from its earlier response
to include memory subsumption) and produced a proof:
\begin{leancode*}{linenos=false}
theorem eval_exists_answer (heval : Eval C m e Q) :
    ∃ m' e', e'.IsAns ∧ m'.subsumes m ∧ Q e' m'
\end{leancode*}
The proof proceeds by induction on the evaluation derivation.
Base cases (values, variables, capability invocations) are immediate.
Sequencing cases (\texttt{letin}, \texttt{unpack}, \texttt{cond})
apply the induction hypothesis to obtain an intermediate answer,
then use the continuation to reach the final answer.
The agent completed this in under 5 minutes
with 21 tool calls (6 edits, 5 compiler invocations).

This example shows how agents can formalize informal conjectures.
The human had an intuition about what should hold.
The agent made this intuition precise and verified it with a formal proof.

\subsubsection{Other Examples}
There are other kinds of interactions we observed, but they occurred less often.
One example is idealization\footnote{\url{https://apa.univalence.xyz/\#/tasks/fe72733f-3fb3-4a0d-820f-28cd8eaa64d2}},
where the human is thinking through the design of the formal system or the proof strategies,
and they use the agent as a thinking buddy to brainstorm ideas.
In our case study,
we found that 
although the agent followed the conversation well,
it is unable to contribute novel insights and always blindly acknowledges the human's ideas.
This exhibits the current limitation of agents in creative reasoning tasks:
it is an excellent proof engineer, but not an original theorist.
Another example is chores and maintenance,
where the human delegates tedious routine tasks,
like adding documentation comments\footnote{\url{https://apa.univalence.xyz/\#/tasks/8efffdc3-91f0-45dd-b523-fe1ef496c79c}}, 
module structure refactoring\footnote{\url{https://apa.univalence.xyz/\#/tasks/53af846a-0e1e-49e4-8499-0af57bf266db}},
and style consistency fixes\footnote{\url{https://apa.univalence.xyz/\#/tasks/1e4e9d0a-c3e3-4e25-b0b8-e0b323fc0982}},
to the agent.
The agent performed these tasks well:
it is a reliable copilot for mundane maintenance work
that is orthogonal to the mathematical content of the mechanization.

\subsection{Patterns and Limitations}

Through these interactions in our case study,
we observed recurring patterns in how humans guide agents
and where agents succeed or struggle.
The interactions reveal a natural division:
humans provide structure, context, and mathematical insight,
while agents handle the mechanical work of proof engineering.
We discuss the guidance patterns that emerged
and the limitations we encountered.

\subsubsection{Human Guidance Patterns}
Human guidance operates at two levels: system-level and task-level.

\emph{System-level instructions} persist across sessions
in configuration files (conventionally named \texttt{CLAUDE.md} or \texttt{AGENTS.md}).
These encode domain knowledge the agent should always apply:
tool preferences (``use \texttt{lean4check} rather than \texttt{lake build}''),
tactic idioms (``use the \texttt{calc} mode for equalities''),
debugging techniques (``use \texttt{trace\_state} to inspect goal contexts''),
and corrections for tactics with non-obvious semantics
(e.g., documenting that \texttt{rename\_i} names variables
from the bottom of the goal context, not the top).

\emph{Task-level prompts} guide individual interactions.

These fall into several categories:
\begin{itemize}
\item \emph{Directive prompts} state the goal directly:
  ``prove theorem \texttt{step\_preserves\_eval}''
  or ``fix all errors in \texttt{Semantics/Props.lean}.''
\item \emph{Strategic prompts} suggest proof approaches:
  ``use induction on the evaluation derivation''
  or ``apply the monotonicity lemma.''
\item \emph{Incremental prompts} decompose work into stages:
  ``first state the theorem and sketch the cases, leaving \texttt{sorry} placeholders''
  or ``let's proceed step by step: start by exploring related definitions.''
\item \emph{Onboarding prompts} orient the agent within the codebase:
  ``read the module documentation to understand the context''
  or ``explore related files before attempting the proof.''
\item \emph{Autonomous prompts} delegate extended work:
  ``continue fixing errors until the module compiles''
  or ``work through the remaining cases while I step away.''
\end{itemize}
Concrete examples of these prompt types can be found in the examples of Section~\ref{sec:examples}.

\subsubsection{Agent Limitations}

In our case study,
agents exhibit some recurring patterns of limitations.
We discuss each of them below.

\emph{Subtle proof state handling.}
Agents struggle with prover-specific idioms that have non-obvious semantics.
A recurring example\footnote{\url{https://apa.univalence.xyz/\#/tasks/e69d9c6d-759a-413e-8927-43848ac726f0}} is the \texttt{rename\_i} tactic in Lean 4,
which names anonymous hypotheses (marked with \texttt{\dag}) in the goal context.
The tactic names variables starting from the \emph{bottom} of the context,
which is counter-intuitive: \texttt{rename\_i x y z} names the last three
anonymous variables, not the first three.
Agents repeatedly misuse this tactic 
despite multiple corrections and system-level instructions.

\emph{Context retrieval failures.}
Agents sometimes fail to locate relevant lemmas that exist in the codebase.
This happens when lemma names do not obviously match the goal,
or when the lemma resides in an unexpected module.
The agent then attempts to re-prove something that already exists,
or simply gets stuck.
Strategic prompts that name specific lemmas resolve this,
but the failure reveals limits in codebase comprehension.

\emph{Long-term planning.}
Complex proofs requiring multi-step decomposition challenge agents.
When a proof requires introducing non-trivial auxiliary lemmas,
agents often get lost without human guidance.
Incremental prompts that break the proof into stages help,
but the agent often fails to initiate such decomposition itself.

\emph{Creative reasoning.}
As noted in Section~\ref{sec:examples},
agents fall short of contributing novel mathematical insights.
When used as a thinking partner for design ideation,
agents follow the conversation and acknowledge ideas,
but they are unable to propose original and reasonable ideas.
They are capable proof engineers but not original theorists.

\subsection{Discussion}

What we observed
suggests two emerging workflows:
rapid prototyping of formalizations,
and potential natural language interfaces of mechanized proofs.

\paragraph{Fast and Reliable Prototyping of Formalizations.}
Developing a mechanized formalization typically involves an iterative cycle:
update a definition, repair the proofs that break,
discover a problem in the original design, and repeat.
The bottleneck is proof repair.
A single definition change can invalidate dozens of lemmas,
and manually fixing them in the proof assistant is tedious work
that delays feedback on whether the design is sound.

Agentic proof automation compresses this cycle.
The human updates a definition;
the agent repairs the affected proofs;
if repairs succeed, the design is validated;
if not, the agent often identifies why
(as in the \texttt{step\_masked} example in Section~\ref{sec:proof-repair}, 
where the agent diagnosed a flaw in the semantics rather than in the proof).
Either way, feedback arrives in minutes rather than hours.
This rapid iteration enables a style of development
where formal verification becomes part of the design process,
not a burden imposed after the fact.

\paragraph{Natural Language Interface for Mechanized Proofs.}
Throughout the case study,
we observed that most communications from the human to the agent are
in natural language: stating goals, sketching strategies, explaining obstacles.
Agents translated these informal descriptions into formal proof scripts
and reported results back in natural language.
This pattern suggests a possibility where
humans interact with machine-verified proofs
in pure natural language,
with agents serving as the translation layer
between human mathematical intuition and machine-checked proofs.

Currently, 
human intervention remains necessary,
which requires expertise in the proof assistant from the human.
But as models improve,
this gap may narrow.
A mathematician might eventually state a theorem in plain natural language,
provide a rough proof idea,
and receive a fully mechanized proof in return.
When there are flaws in the reasoning,
the agent can also point them out and even explain counterexamples
in natural language.
 
\section{Quantitative Evaluation}

In this section,
we present a quantitative analysis of the interactions between the human and the LLM agent
to characterize the performance patterns of agentic proof automation.

\subsection{Data Collection and Annotation}

The raw data consists of conversation histories between the human and the agent.
Each conversation forms a \emph{session} containing multiple messages from both sides.
After filtering out empty and trivial interactions (e.g., messages for testing connectivity or sessions without proof-related tool calls),
we obtain 58 sessions with 513 messages in total.

We then manually group consecutive messages into \emph{tasks}.
A task represents a single unit of work: 
a request from the human followed by one or more agent responses until completion or abandonment.
One session may contain multiple tasks when the human issues several requests.

Each task is annotated with a \emph{category} of the work involved:
\begin{itemize}
  \item \textbf{Proof}: the human states a theorem; the agent proves it.
  \item \textbf{Repair}: the human makes changes that break existing proofs; the agent fixes them.
  \item \textbf{Refactor}: update definitions and proofs based on natural language instructions.
  \item \textbf{State and Prove}: the human gives a natural language description of a property; the agent states and proves it.
  \item \textbf{Query}: answer questions about the codebase or Lean concepts.
  \item \textbf{Chore}: routine tasks such as fixing style warnings or adding imports.
\end{itemize}

Each task is also annotated with an \emph{outcome}:
\begin{itemize}
  \item \textbf{Success}: the agent completes the task without human intervention.
  \item \textbf{Success with intervention}: the agent completes the task after human guidance (natural language hints, code edits, or both).
  \item \textbf{Partial}: the agent makes progress but leaves some proof obligations (\texttt{sorry}s) unfilled.
  \item \textbf{Failure}: the agent fails to complete the task even with human intervention.
\item \textbf{Problem identified}: the agent identifies an issue in the specification or reasoning, and escalates it to the human.
\end{itemize}

In total, we annotated 189 tasks from the 58 sessions.
All raw sessions and annotated tasks are available in the online explorer.

\subsection{Results}

\begin{table}[t]
\centering
\caption{Quantitative evaluation of agentic proof automation tasks.
  Success includes tasks completed with human intervention.
  Checks and Lines show mean number of \texttt{lean4check} invocations and lines changed.
  Intervention shows percentage of tasks requiring human guidance.}
\label{tab:evaluation}
\small
\begin{tabular}{lrrrrrrr}
\toprule
Category & Total & Success & Partial & Fail & Checks & Lines & Intervention \\
\midrule
Proof & $51$ & $41$ & $6$ & $4$ & $8.3$ & $82.6$ & $20$\% \\
Repair & $48$ & $43$ & $2$ & $1$ & $4.9$ & $40.8$ & $8$\% \\
Refactor & $35$ & $28$ & $4$ & $3$ & $6.6$ & $61.8$ & $23$\% \\
State+Prove & $23$ & $20$ & $2$ & $1$ & $8.3$ & $129.3$ & $35$\% \\
Query & $21$ & $21$ & $0$ & $0$ & $0.3$ & $0.1$ & $0$\% \\
Chore & $11$ & $11$ & $0$ & $0$ & $2.1$ & $5.9$ & $9$\% \\
\midrule
Total & $189$ & $164$ & $14$ & $9$ & $5.9$ & $60.2$ & $16$\% \\
\bottomrule
\end{tabular}
\end{table}

Table~\ref{tab:evaluation} summarizes the results.
The agent completes 164 of 189 tasks (87\%), with only 9 failures (5\%) and 16\% requiring human intervention.
Performance varies by category:
Query and Chore achieve 100\% success as they involve no proof synthesis,
while the four proof-related categories show lower success rates.

\paragraph{The value of existing context.}
Among proof-related tasks, the agent performs the best for Repair tasks:
90\% success rate (43/48), only 8\% intervention, and fewer iterations (4.9 checks vs.\ 8.3 for Proof).
The existing broken proof provides valuable context.
The agent can see the intended approach and often needs only local fixes rather than synthesizing a proof from scratch.
In contrast, Proof tasks start from a bare theorem statement,
requiring the agent to discover the proof strategy independently.

\paragraph{Formalization is hard.}
State and Prove tasks have the highest intervention rate (35\%) and produce the most code (129 lines on average).
These tasks require the agent to translate informal natural language descriptions into formal theorem statements before proving them.
This formalization step introduces ambiguity:
the human may need to clarify the intended specification or correct misinterpretations.
Once the statement is correct, the proof effort is comparable to Proof tasks (both average 8.3 checks).

\paragraph{Iteration enables success.}
The number of \texttt{lean4check} invocations reflects task difficulty.
Proof and State and Prove tasks require the most iterations (8.3),
Refactor and Repair need fewer (6.6 and 4.9),
and Query tasks need almost none (0.3).
The feedback loop between the agent and the type checker is essential:
the agent proposes proof steps, receives error messages, and refines its approach.
More complex tasks require more iterations.
 
\section{Related Work}

Three research areas are related to our case study:
classical proof automation and proof repair, 
neural theorem proving, 
and autonomous coding agents. 
We survey each area in the following sections.

\subsection{Proof Automation and Proof Repair}
The automation of interactive theorem provers (ITPs) 
has traditionally relied on symbolic search algorithms and heuristic methods. 
In the Isabelle/HOL ecosystem, Sledgehammer~\cite{BohmeN10} represents the state-of-the-art in ``hammer'' systems. 
It translates higher-order logic goals into first-order logic problems and dispatches them 
to external Automated Theorem Provers (ATPs).
If an external solver finds a proof,
Sledgehammer reconstructs it within Isabelle's trusted kernel. 
In Coq,
tools like Tactician~\cite{BlaauwbroekUG20} have focused on learning from user-written tactic scripts.
Tactician employs $k$-nearest neighbor ($k$-NN) algorithms to predict the next tactic 
based on the current proof state, 
retrieving knowledge from existing libraries. 
Unlike external hammers, it operates directly within the ITP's tactic language.
For Lean 4,
Aesop~\cite{LimpergF23} (Automated Extensible Search for Obvious Proofs)
provides a white-box automation framework.
Aesop conducts a best-first tree search over a user-configurable set of rules.
While highly effective for discharging routine goals,
Aesop is fundamentally a search algorithm driven
by fixed heuristics.

Related to proof automation is \emph{proof repair}:
adapting proofs when specifications change.
\textsc{Pumpkin Patch}~\cite{ringer2018adapting} pioneered example-based patching,
later generalized by \textsc{Pumpkin Pi}~\cite{ringer2021proof},
which transports proofs across type equivalences to support complex refactorings
including algebraic ornaments~\cite{ringer2019ornaments}
and quotient types~\cite{viola2025proof}.
\textsc{Sisyphus}~\cite{gopinathan2023mostly} complements these static approaches
with dynamic analysis for synthesizing loop invariants.
These tools excel at structured, predictable changes
but lack the flexibility to handle arbitrary modifications
or provide natural language explanations of failures.

\subsection{Interactive Theorem Proving with LLMs}
The integration of LLMs with ITPs,
often termed Neural Theorem Proving (NTP),
has evolved rapidly.
Early work such as GPT-f~\cite{PoluS20} showed that
Transformers could generate valid proof steps when trained on formalized libraries.
Llemma~\cite{AzerbayevSPSDMJDBW23} improved on this by pre-training on mathematical code.
LeanDojo~\cite{YangSGCSYGPA23} enabled fine-grained interaction with Lean,
and ReProver~\cite{YangSGCSYGPA23} introduced retrieval-augmented generation
to select relevant premises from Mathlib.

Recent approaches have moved beyond static next-step prediction
to agentic workflows that learn from interaction.
Copra~\cite{ThakurWC23} uses GPT-4 within a stateful backtracking search,
reading error messages to prune incorrect paths.
DeepSeek-Prover~\cite{XinRSSSWLLZDGZYGWLR24} and InternLM-StepProver~\cite{WuHZYWLC24}
combine reinforcement learning with Monte-Carlo Tree Search.
AlphaProof~\cite{AlphaProof24} achieves Olympiad-level performance
by training on millions of synthetic problems.
Other systems focus on human-in-the-loop or continuous learning workflows:
Lean Copilot~\cite{SongYA24} suggests tactics interactively,
and LeanAgent~\cite{KumarappanTSGXA24} uses curriculum learning
to progressively tackle harder theorems.

Our case study differs from these systems in scope and setup.
Most NTP research targets mathematical benchmarks
where each problem is a self-contained theorem statement.
Our work focuses on in-context proof engineering
within a large, evolving formal development
where proofs depend on surrounding definitions, lemmas, and design decisions.
Furthermore, we use no post-training:
our setup relies entirely on off-the-shelf LLM agents
with a single tool for compiler feedback.
This simplicity makes the workflow immediately accessible to practitioners
and allows it to improve as frontier models advance.

\subsection{Agentic LLMs for Software Engineering}
Our case study also connects to advancements in autonomous software engineering agents, which share the paradigm of exploring a structured environment (a codebase) to achieve a goal.

SWE-agent~\cite{YangJWLYNP24} demonstrated
that simplifying commands for file navigation and editing
significantly improves an LLM's ability to resolve GitHub issues autonomously.
Closer to our domain,
ProofWright~\cite{ChatterjeeEtAl25} applies agentic workflows to 
the formal verification of CUDA kernels, 
generating both code and correctness proofs. 
However, while ProofWright focuses on high-performance kernels, 
our case study investigates the application of general-purpose coding agents 
to the mechanization of complex metatheory in programming language semantics.
 
\section{Conclusion}

We presented a case study of agentic proof automation:
mechanizing System Capless's semantic type soundness in over 14,000 lines of Lean 4
using off-the-shelf LLM agents with a simple proof checker integration.
The quantitative evaluation across 189 tasks shows an 87\% success rate,
with only 16\% requiring human intervention.
The case study reveals a natural division of labor:
humans provide mathematical insight (definitions, theorems, proof strategies)
while agents handle proof engineering through iterative refinement against compiler feedback.
All 58 sessions and 189 annotated tasks are available through an interactive explorer,
allowing readers to examine the raw interactions and see how guidance patterns translate into proof attempts.
The case study mechanization is also open-sourced for experimentation and extension.

The agentic proof automation scheme illustrated by our case study
generalizes beyond Lean and scales with frontier model improvements.
As agents become more capable,
the barrier to mechanized proof drops,
making machine-checked proof development more accessible to
practitioners and theorists who lack years of theorem prover expertise.

Our case study
examines a single formal development in Lean 4.
A natural direction for future work is a standardized benchmark for agentic proof automation.
Unlike existing benchmarks that evaluate isolated theorem proving,
such a benchmark would evaluate in-context proof engineering within evolving codebases.

\bibliographystyle{splncs04}
\bibliography{references}

@article{Xu2025WhatsInTheBox,
  author    = {Yichen Xu and Oliver Bračevac and Cao Nguyen Pham and Martin Odersky},
  title     = {What’s in the Box: Ergonomic and Expressive Capture Tracking over Generic Data Structures},
  journal   = {Proceedings of the ACM on Programming Languages},
  volume    = {9},
  number    = {OOPSLA2},
  pages     = {1726--1753},
  year      = {2025},
  doi       = {10.1145/3763112},
  publisher = {Association for Computing Machinery},
  month     = oct,
  url       = {https://dl.acm.org/doi/10.1145/3763112}
}

@manual{Coq2020,
  author    = {The Coq Development Team},
  title     = {The Coq Proof Assistant, Reference Manual},
  organization = {INRIA, CNRS, École Polytechnique},
  year      = {2020},
  url       = {https://coq.inria.fr},
  note      = {Version 8.12}
}

@inproceedings{lean2017,
  author    = {Leonardo de Moura and Soonho Kong and Jeremy Avigad and Floris van Doorn and Jakob von Raumer},
  title     = {The Lean Theorem Prover (System Description)},
  booktitle = {Automated Deduction – CADE 26},
  editor    = {Nicolas Ollinger and Hélène Kirchner},
  year      = {2017},
  pages     = {378--388},
  publisher = {Springer},
  doi       = {10.1007/978-3-319-61407-3_26}
}

@book{isabelle-hol-2002,
  author    = {Tobias Nipkow and Lawrence C. Paulson and Markus Wenzel},
  title     = {Isabelle/HOL: A Proof Assistant for Higher-Order Logic},
  year      = {2002},
  publisher = {Springer},
  series    = {Lecture Notes in Computer Science},
  volume    = {2283},
  isbn      = {3-540-43376-7},
  doi       = {10.1007/3-540-45949-9}
}

@article{leroy2008compcert,
  author       = {Xavier Leroy},
  title        = {Formal Verification of a Realistic Compiler},
  journal      = {Communications of the ACM},
  year         = {2008},
  volume       = {51},
  number       = {7},
  pages        = {107--115},
  doi          = {10.1145/1378704.1378709},
  url          = {https://doi.org/10.1145/1378704.1378709}
}

@inproceedings{Boehme2010Sledgehammer,
  author    = {Sascha Böhme and Tobias Nipkow},
  title     = {Sledgehammer: Judgement Day},
  booktitle = {International Joint Conference on Automated Reasoning (IJCAR)},
  series    = {LNCS},
  volume    = {6173},
  pages     = {107--121},
  publisher = {Springer},
  year      = {2010},
  doi       = {10.1007/978-3-642-14203-1_9}
}

@article{Blanchette2013SledgehammerSMT,
  author    = {Jasmin Christian Blanchette and Sascha Böhme and Lawrence C. Paulson},
  title     = {Extending {Sledgehammer} with {SMT} Solvers},
  journal   = {Journal of Automated Reasoning},
  volume    = {51},
  number    = {1},
  pages     = {109--128},
  year      = {2013},
  doi       = {10.1007/s10817-013-9278-5}
}

@article{Gauthier2021TacticToe,
  author    = {Thibault Gauthier and Cezary Kaliszyk and Josef Urban},
  title     = {{TacticToe}: Learning to Prove with Tactics},
  journal   = {Journal of Automated Reasoning},
  volume    = {65},
  number    = {2},
  pages     = {257--286},
  year      = {2021},
  doi       = {10.1007/s10817-020-09580-x}
}

@inproceedings{Blaauwbroek2020Tactician,
  author    = {Lasse Blaauwbroek and Josef Urban and Herman Geuvers},
  title     = {The {Tactician}: A Seamless, Interactive Tactic Learner and Prover for {Coq}},
  booktitle = {Intelligent Computer Mathematics (CICM)},
  series    = {LNCS},
  volume    = {12236},
  pages     = {271--277},
  publisher = {Springer},
  year      = {2020},
  doi       = {10.1007/978-3-030-53518-6_17}
}

@inproceedings{Limperg2023Aesop,
  author    = {Jannis Limperg and Asta Halkjær From},
  title     = {{Aesop}: White-Box Best-First Proof Search for {Lean}},
  booktitle = {Certified Programs and Proofs (CPP)},
  pages     = {253--266},
  publisher = {ACM},
  year      = {2023},
  doi       = {10.1145/3573105.3575671}
}

@inproceedings{Hamza2019Stainless,
  author    = {Jad Hamza and Nicolas Voirol and Viktor Kuncak},
  title     = {System {FR}: Formalized Foundations for the {Stainless} Verifier},
  booktitle = {Proceedings of the ACM on Programming Languages, OOPSLA},
  volume    = {3},
  number    = {OOPSLA},
  pages     = {148:1--148:30},
  year      = {2019},
  doi       = {10.1145/3360592}
}

@inproceedings{Vazou2014LiquidHaskell,
  author    = {Niki Vazou and Ranjit Jhala and David Van Horn and Soonho Kong},
  title     = {{LiquidHaskell}: Experience with Refinement Types in the Real World},
  booktitle = {ACM SIGPLAN International Conference on Functional Programming (ICFP)},
  pages     = {39--51},
  year      = {2014},
  publisher = {ACM},
  doi       = {10.1145/2633357.2633366}
}

@article{boruchgruszecki_capturing_2023,
title = {{Capturing Types}},
author = {Boruch-Gruszecki, Aleksander and Odersky, Martin and Lee, Edward and Lhoták, Ondřej and Brachthäuser, Jonathan},
year = {2023},
journal = {{ACM Trans. Program. Lang. Syst.}},
number = {4},
doi = {10.1145/3618003},
issn = {0164-0925},
pages = {21:1--21:52},
url = {https://dl.acm.org/doi/10.1145/3618003},
volume = {45}
}

@article{timany2024logical-type-soundness,
author = {Timany, Amin and Krebbers, Robbert and Dreyer, Derek and Birkedal, Lars},
title = {A Logical Approach to Type Soundness},
journal = {J. ACM},
year = {2024},
volume = {71},
number = {6},
articleno = {40},
numpages = {75},
month = nov,
publisher = {Association for Computing Machinery},
address = {New York, NY, USA},
issn = {0004-5411},
doi = {10.1145/3676954},
url = {https://doi.org/10.1145/3676954}
}

@inproceedings{BohmeN10,
  author       = {Sascha B{\"o}hme and Tobias Nipkow},
  title        = {Sledgehammer: Judgement Day},
  booktitle    = {Automated Reasoning - 5th International Joint Conference, {IJCAR}
                  2010, Edinburgh, UK, July 16-19, 2010. Proceedings},
  series       = {Lecture Notes in Computer Science},
  volume       = {6173},
  pages        = {107--121},
  publisher    = {Springer},
  year         = {2010},
  doi          = {10.1007/978-3-642-14203-1_9},
}

@inproceedings{BlaauwbroekUG20,
  author       = {Lasse Blaauwbroek and Josef Urban and Herman Geuvers},
  title        = {The Tactician - {A} Seamless, Interactive Tactic Learner and Prover
                  for Coq},
  booktitle    = {Intelligent Computer Mathematics - 13th International Conference,
                  {CICM} 2020, Bertinoro, Italy, July 26-31, 2020, Proceedings},
  series       = {Lecture Notes in Computer Science},
  volume       = {12236},
  pages        = {271--277},
  publisher    = {Springer},
  year         = {2020},
  doi          = {10.1007/978-3-030-53518-6_17},
}

@inproceedings{LimpergF23,
  author       = {Jannis Limperg and Asta Halkj{\ae}r From},
  title        = {Aesop: White-Box Best-First Proof Search for Lean},
  booktitle    = {Proceedings of the 12th {ACM} {SIGPLAN} International Conference
                  on Certified Programs and Proofs, {CPP} 2023, Boston, MA, USA,
                  January 16-17, 2023},
  pages        = {253--266},
  publisher    = {{ACM}},
  year         = {2023},
  doi          = {10.1145/3573105.3575671},
}

@article{PoluS20,
  author       = {Stanislas Polu and Ilya Sutskever},
  title        = {Generative Language Modeling for Automated Theorem Proving},
  journal      = {CoRR},
  volume       = {abs/2009.03393},
  year         = {2020},
  url          = {https://arxiv.org/abs/2009.03393},
  eprinttype   = {arXiv},
  eprint       = {2009.03393},
}

@article{AzerbayevSPSDMJDBW23,
  author       = {Zhangir Azerbayev and Hailey Schoelkopf and Keiran Paster and
                  Marco Dos Santos and Stephen McAleer and Albert Q. Jiang and
                  Jia Deng and Stella Biderman and Sean Welleck},
  title        = {Llemma: An Open Language Model for Mathematics},
  journal      = {CoRR},
  volume       = {abs/2310.10631},
  year         = {2023},
  url          = {https://arxiv.org/abs/2310.10631},
  eprinttype   = {arXiv},
  eprint       = {2310.10631},
}

@inproceedings{YangSGCSYGPA23,
  author       = {Kaiyu Yang and Aidan M. Swope and Alex Gu and Rahul Chalamala and
                  Peiyang Song and Shixing Yu and Saber Godber and Ryan Prenger and
                  Anima Anandkumar},
  title        = {LeanDojo: Theorem Proving with Retrieval-Augmented Language Models},
  booktitle    = {Advances in Neural Information Processing Systems 36: Annual
                  Conference on Neural Information Processing Systems 2023,
                  NeurIPS 2023, New Orleans, LA, USA, December 10-16, 2023},
  year         = {2023},
  url          = {https://arxiv.org/abs/2306.15626},
}

@article{ThakurWC23,
  author       = {Amitayush Thakur and Yeming Wen and Swarat Chaudhuri},
  title        = {{COPRA}: In-Context Learning for Formal Theorem Proving},
  journal      = {CoRR},
  volume       = {abs/2310.04353},
  year         = {2023},
  url          = {https://arxiv.org/abs/2310.04353},
  eprinttype   = {arXiv},
  eprint       = {2310.04353},
}

@article{XinRSSSWLLZDGZYGWLR24,
  author       = {Huajian Xin and Z. Z. Ren and Junxiao Song and Zhihong Shao and
                  Wanjia Zhao and Haocheng Wang and Bo Liu and Liyue Zhang and
                  Xuan Lu and Qiushi Du and Wenjun Gao and Qihao Zhu and Dejian Yang
                  and Zhibin Gou and Z. F. Wu and Fuli Luo and Chong Ruan},
  title        = {DeepSeek-Prover-V1.5: Harnessing Proof Assistant Feedback for
                  Reinforcement Learning and Monte-Carlo Tree Search},
  journal      = {CoRR},
  volume       = {abs/2408.08152},
  year         = {2024},
  url          = {https://arxiv.org/abs/2408.08152},
  eprinttype   = {arXiv},
  eprint       = {2408.08152},
}

@article{WuHZYWLC24,
  author       = {Zijian Wu and Suozhi Huang and Zhejian Zhou and Huaiyuan Ying and
                  Jiayu Wang and Dahua Lin and Kai Chen},
  title        = {InternLM2.5-StepProver: Advancing Automated Theorem Proving via
                  Expert Iteration on Large-Scale {LEAN} Problems},
  journal      = {CoRR},
  volume       = {abs/2410.15700},
  year         = {2024},
  url          = {https://arxiv.org/abs/2410.15700},
  eprinttype   = {arXiv},
  eprint       = {2410.15700},
}

@misc{AlphaProof24,
  author       = {{AlphaProof Team} and {AlphaGeometry Team}},
  title        = {{AI} achieves silver-medal standard solving International Mathematical
                  Olympiad problems},
  howpublished = {Google DeepMind Blog},
  year         = {2024},
  url          = {https://deepmind.google/discover/blog/ai-solves-imo-problems-at-silver-medal-level/},
  note         = {Published July 25, 2024},
}

@article{SongYA24,
  author       = {Peiyang Song and Kaiyu Yang and Anima Anandkumar},
  title        = {Towards Large Language Models as Copilots for Theorem Proving in Lean},
  journal      = {CoRR},
  volume       = {abs/2404.12534},
  year         = {2024},
  url          = {https://arxiv.org/abs/2404.12534},
  eprinttype   = {arXiv},
  eprint       = {2404.12534},
}

@article{KumarappanTSGXA24,
  author       = {Adarsh Kumarappan and Mo Tiwari and Peiyang Song and
                  Robert Joseph George and Chaowei Xiao and Anima Anandkumar},
  title        = {LeanAgent: Lifelong Learning for Formal Theorem Proving},
  journal      = {CoRR},
  volume       = {abs/2410.06209},
  year         = {2024},
  url          = {https://arxiv.org/abs/2410.06209},
  eprinttype   = {arXiv},
  eprint       = {2410.06209},
}

@inproceedings{YangJWLYNP24,
  author       = {John Yang and Carlos E. Jimenez and Alexander Wettig and
                  Kilian Lieret and Shunyu Yao and Karthik Narasimhan and Ofir Press},
  title        = {{SWE-agent}: Agent-Computer Interfaces Enable Automated Software
                  Engineering},
  booktitle    = {Advances in Neural Information Processing Systems 37: Annual
                  Conference on Neural Information Processing Systems 2024,
                  NeurIPS 2024, Vancouver, BC, Canada, December 10-15, 2024},
  year         = {2024},
  url          = {https://arxiv.org/abs/2405.15793},
}

@article{ChatterjeeEtAl25,
  author       = {Bodhisatwa Chatterjee and others},
  title        = {ProofWright: Towards Agentic Formal Verification of {CUDA}},
  journal      = {CoRR},
  volume       = {abs/2511.12294},
  year         = {2025},
  url          = {https://arxiv.org/abs/2511.12294},
  eprinttype   = {arXiv},
  eprint       = {2511.12294},
}

@article{vaswani2017attention,
  title={Attention is all you need},
  author={Vaswani, Ashish and Shazeer, Noam and Parmar, Niki and Uszkoreit, Jakob and Jones, Llion and Gomez, Aidan N and Kaiser, {\L}ukasz and Polosukhin, Illia},
  journal={Advances in neural information processing systems},
  volume={30},
  year={2017}
}

@article{devlin2018bert,
  title={Bert: Pre-training of deep bidirectional transformers for language understanding},
  author={Devlin, Jacob and Chang, Ming-Wei and Lee, Kenton and Toutanova, Kristina},
  journal={arXiv preprint arXiv:1810.04805},
  year={2018}
}

@inproceedings{brown2020language,
  title={Language models are few-shot learners},
  author={Brown, Tom and Mann, Benjamin and Ryder, Nick and Subbiah, Melanie and Kaplan, Jared D and Dhariwal, Prafulla and Neelakantan, Arvind and Shyam, Pranav and Sastry, Girish and Askell, Amanda and others},
  booktitle={Advances in neural information processing systems},
  volume={33},
  pages={1877--1901},
  year={2020}
}

@article{radford2019language,
  title={Language models are unsupervised multitask learners},
  author={Radford, Alec and Wu, Jeffrey and Child, Rewon and Luan, David and Amodei, Dario and Sutskever, Ilya},
  journal={OpenAI blog},
  volume={1},
  number={4},
  pages={9},
  year={2019}
}

@misc{touvron2023llama,
  title={LLaMA: Open and Efficient Foundation Language Models},
  author={Touvron, Hugo and Lavril, Thibaut and Izacard, Gautier and Martinet, Xavier and Lachaux, Marie-Anne and Lacroix, Timoth{\'e}e and Rozi{\`e}re, Baptiste and Goyal, Naman and Hambro, Eric and Zoph, B{\'e}la and others},
  year={2023},
  eprint={2302.13971},
  archivePrefix={arXiv},
  primaryClass={cs.CL}
}

@misc{chowdhery2022palm,
  title={PaLM: Scaling Language Modeling with Pathways},
  author={Chowdhery, Aakanksha and Narang, Sharan and Devlin, Jacob and Bosma, Maarten and Mishra, Gaurav and Roberts, Adam and Barham, Paul and Chung, Hyung Won and Sutton, Charles and Gehrmann, Sebastian and others},
  year={2022},
  eprint={2204.02311},
  archivePrefix={arXiv},
  primaryClass={cs.CL}
}

@inproceedings{ringer2018adapting,
  author    = {Ringer, Talia and Yazdani, Nathaniel and Leo, John and Grossman, Dan},
  title     = {Adapting Proof Automation to Adapt Proofs},
  booktitle = {Proceedings of the 7th ACM SIGPLAN International Conference on Certified Programs and Proofs (CPP)},
  year      = {2018},
  pages     = {115--129},
  publisher = {ACM},
  doi       = {10.1145/3167080}
}

@inproceedings{ringer2021proof,
  author    = {Ringer, Talia and Porter, RanDair and Yazdani, Nathaniel and Leo, John and Grossman, Dan},
  title     = {Proof Repair Across Type Equivalences},
  booktitle = {Proceedings of the 42nd ACM SIGPLAN International Conference on Programming Language Design and Implementation (PLDI)},
  year      = {2021},
  pages     = {112--127},
  publisher = {ACM},
  doi       = {10.1145/3453483.3454033}
}

@inproceedings{ringer2019ornaments,
  author    = {Ringer, Talia and Yazdani, Nathaniel and Leo, John and Grossman, Dan},
  title     = {Ornaments for Proof Reuse in Coq},
  booktitle = {10th International Conference on Interactive Theorem Proving (ITP 2019)},
  year      = {2019},
  series    = {LIPIcs},
  volume    = {141},
  pages     = {26:1--26:19},
  doi       = {10.4230/LIPIcs.ITP.2019.26}
}

@article{viola2025proof,
  author    = {Viola, Cosmo and Fan, Max and Ringer, Talia},
  title     = {Proof Repair Across Quotient Type Equivalences},
  journal   = {Proc. ACM Program. Lang.},
  volume    = {9},
  number    = {OOPSLA2},
  year      = {2025},
  publisher = {ACM},
  note      = {To appear}
}

@inproceedings{gopinathan2023mostly,
  author    = {Gopinathan, Kiran and Keoliya, Mayank and Sergey, Ilya},
  title     = {Mostly Automated Proof Repair for Verified Libraries},
  booktitle = {Proceedings of the 44th ACM SIGPLAN International Conference on Programming Language Design and Implementation (PLDI)},
  year      = {2023},
  pages     = {25--49},
  publisher = {ACM},
  doi       = {10.1145/3591239}
}

\clearpage
\appendix

\section{Formal Definitions of System Capless}
\label{sec:appendix-capless}

This appendix presents the formal definitions of System Capless,
the target system formalized in Lean 4 as the case study of this paper.
System Capless is a core calculus for capture checking,
extending System $\textsf{CC}_{<:\Box}$~\cite{boruchgruszecki_capturing_2023}
with capture polymorphism.

\subsection{Syntax}
\label{sec:appendix-syntax}

\begin{figure}[htbp]
\begin{flushleft}\small
\noindent
{\footnotesize\begin{multicols}{3}\noindent
\begin{flalign*}
  x,\,y,\,z         \tag*{\textbf{Variable}}\\
  X                 \tag*{\textbf{Type Variable}}\\
  c                 \tag*{\textbf{Capture Variable}}\\
  s,\,t,\,u\coloneqq\ &           \tag*{\textbf{Term}}\\
  &a                              \tag*{answer}\\
  &x\,y                              \tag*{app.}\\
  &x[S]                              \tag*{type app.}\\
  &x[c]                              \tag*{capture app.}\\
  &\textsf{let}\ x = t \ \textsf{in}\ u                  \tag*{let}\\
  &\textsf{let}\ \langle c, x \rangle = t \ \textsf{in}\ u    \tag*{existential let}\\
  v\coloneqq\ &           \tag*{\textbf{Value}}\\
  &\lambda(x: T)t                      \tag*{term lambda}\\
\end{flalign*}
\begin{flalign*}
  &\lambda[X<:S]t                      \tag*{type lambda}\\
  &\lambda[c<:B]t                    \tag*{capt. lambda}\\
  &\langle C, x\rangle                    \tag*{pack}\\
  E,\,F\coloneqq\ &           \tag*{\textbf{Existential Type}}\\
  &\exists c.\, T   \tag*{existential}\\
  &T   \tag*{type}\\
  R,\,S\coloneqq\ &           \tag*{\textbf{Shape Type}}\\
  &\top                    \tag*{top}\\
  &X                    \tag*{type variable}\\
  &\forall(x: T)E             \tag*{term function}\\
  &\forall[X<:S]E             \tag*{type function}\\
  &\forall[c<:B]E             \tag*{capt. function}\\
\end{flalign*}
\begin{flalign*}
  a\coloneqq\ & x \mid v           \tag*{\textbf{Answer}}\\
  \theta\coloneqq\ & x \mid c           \tag*{\textbf{Capture}}\\
  C,\,D\coloneqq\ &\{\theta_1,\cdots,\theta_n\}     \tag*{\textbf{Capture Set}}\\
  B\coloneqq\ & * \mid C           \tag*{\textbf{Capture Bound}}\\
  T,\,U\coloneqq\ &           \tag*{\textbf{Type}}\\
  &S\capt C                    \tag*{capturing}\\
  &S                    \tag*{pure}\\
  \Gamma,\,\Delta\coloneqq\ &           \tag*{\textbf{Context}}\\
  &\emptyset                    \tag*{empty}\\
  &\Gamma, x: T                    \tag*{term binding}\\
  &\Gamma, X<:S                    \tag*{type binding}\\
  &\Gamma, c<:B                    \tag*{capt. binding}\\
\end{flalign*}
\end{multicols}}
\vspace{-3em}
\caption{Abstract syntax of System Capless.}\label{fig:syntax}
\end{flushleft}
\end{figure}

Figure~\ref{fig:syntax} shows the abstract syntax of System Capless.
The key extension from System $\textsf{CC}_\textsf{formal}$ is the addition of capture variables $c$,
which enable capture polymorphism through capture abstraction $\lambda[c<:B]t$,
capture application $x[c]$, and existential types $\exists c.\, T$.

\subsection{Type System}
\label{sec:appendix-typing}

\begin{figure*}[htbp]
\footnotesize

\flushleft{\textbf{Typing \quad $C;\Gamma\vdash t : E$}}

\vspace{-1em}

\begin{multicols}{2}

\infrule[\ruledef{var}]
{x: S\capt C\in \Gamma}
{{\{x\}};\Gamma\vdash x : S\capt{\{x\}}}

\infrule[\ruledef{pack}]
{C';\Gamma\vdash x : [c:=C]T}
{{\{\}};\Gamma\vdash\langle C, x\rangle : \exists c.\, T}

\infrule[\ruledef{sub}]
{C;\Gamma\vdash t : E\andalso
 \Gamma\vdash E <: F\\
 \Gamma\vdash C <: C'\andalso
 \Gamma\vdash C',F \ \textsf{{wf}}}
{C';\Gamma\vdash t : F}

\infrule[\ruledef{abs}]
{C;(\Gamma, x: T)\vdash t : E\andalso
 \Gamma\vdash T \ \textsf{{wf}}}
{{\{\}};\Gamma\vdash\lambda(x: T)t : (\forall(x: T) E)\capt \left(C\setminus x\right)}

\infrule[\ruledef{app}]
{C';\Gamma\vdash x : (\forall(z: T) E)\capt C\\
 C';\Gamma\vdash y : T}
{C';\Gamma\vdash x\,y : [z:=y]E}

\infrule[\ruledef{tabs}]
{C;(\Gamma, X<:S)\vdash t : E\andalso\Gamma\vdash S \ \textsf{{wf}}}
{{\{\}};\Gamma\vdash\lambda[X<:S]t : (\forall[X<:S] E)\capt C}

\infrule[\ruledef{tapp}]
{C';\Gamma\vdash x : (\forall[X<:S] E)\capt C}
{C';\Gamma\vdash x[S] : [X:=S]E}

\infrule[\ruledef{cabs}]
{C;(\Gamma, c<:B)\vdash t : E\andalso \Gamma\vdash C \ \textsf{{wf}}}
{{\{\}};\Gamma\vdash\lambda[c<:B]t : (\forall[c<:B]E)\capt C}

\infrule[\ruledef{capp}]
{C';\Gamma\vdash x : (\forall[c<:D]E)\capt C}
{C';\Gamma\vdash x[D] : [c:=D]E}

\infrule[\ruledef{let}]
{C;\Gamma\vdash t : T\andalso
 C;(\Gamma, x: T)\vdash u : E\\
 \Gamma\vdash C, E \ \textsf{{wf}}}
{C;\Gamma\vdash\textsf{let}\ x = t\ \textsf{in}\ u : E}

\infrule[\ruledef{let-e}]
{C;\Gamma\vdash t : \exists c.\, T\\
 C;(\Gamma, c<:*, x: T)\vdash u : F\\
 \Gamma\vdash C, F \ \textsf{{wf}}}
{C;\Gamma\vdash\textsf{let}\ \langle c, x\rangle = t\ \textsf{in}\ u : F}

\end{multicols}

\vspace{-1.5em}
\flushleft{\textbf{Subcapturing \quad $\Gamma\vdash C_1 <: C_2$}}

\begin{multicols}{5}

\infrule[\rruledef{sc-trans}]
{\Gamma\vdash C_1 <: C_2\\
 \Gamma\vdash C_2 <: C_3}
{\Gamma\vdash C_1 <: C_3}

\infrule[\rruledef{sc-var}]
{x : S\capt C \in \Gamma}
{\Gamma\vdash\{x\} <: C}

\infrule[\rruledef{sc-bound}]
{c <: C\in \Gamma}
{\Gamma\vdash\{c\} <: C}

\infrule[\rruledef{sc-elem}]
{C_1\subseteq C_2}
{\Gamma\vdash C_1 <: C_2}

\infrule[\rruledef{sc-set}]
{\Gamma\vdash C_1 <: C\\
 \Gamma\vdash C_2 <: C}
{\Gamma\vdash C_1\cup C_2 <: C}

\end{multicols}

\vspace{-1.6em}
\flushleft{\textbf{Bound Subtyping \quad $\Gamma\vdash B_1 <: B_2$} \text{same as subcapturing plus} $\Gamma\vdash B <: *$}

\flushleft{\textbf{Subtyping \quad $\Gamma\vdash E_1 <: E_2$}}

\vspace{-1em}

\begin{multicols}{4}

\infax[\ruledef{top}]
{\Gamma\vdash S <: \top}

\infax[\ruledef{refl}]
{\Gamma\vdash E <: E}

\infrule[\ruledef{trans}]
{\Gamma\vdash E_1 <: E_2\\ \Gamma\vdash E_2 <: E_3}
{\Gamma\vdash E_1 <: E_3}

\infrule[\ruledef{tvar}]
{\\
 X<:S\in\Gamma}
{\Gamma\vdash X <: S}

\infrule[\ruledef{capt}]
{\Gamma\vdash S_1 <: S_2\\\Gamma\vdash C_1 <: C_2}
{\Gamma\vdash S_1\capt C_1 <: S_2\capt C_2}

\end{multicols}

\begin{multicols}{2}

\infrule[\ruledef{exist}]
{(\Gamma, c<:*)\vdash T_1 <: T_2}
{\Gamma\vdash\exists c.\, T_1 <: \exists c.\, T_2}

\infrule[\ruledef{fun}]
{(\Gamma, x: T_2)\vdash E_1 <: E_2
 \andalso\Gamma\vdash T_2 <: T_1}
{\Gamma\vdash\forall(x: T_1) E_1 <: \forall(x: T_2) E_2}

\infrule[\ruledef{tfun}]
{(\Gamma, X<:S_2)\vdash E_1 <: E_2\andalso\Gamma\vdash S_2 <: S_1}
{\Gamma\vdash\forall[X<:S_1] E_1 <: \forall[X<:S_2] E_2}

\infrule[\ruledef{cfun}]
{(\Gamma, c<:B_2)\vdash E_1 <: E_2\\
 \Gamma\vdash B_2 <: B_1}
{\Gamma\vdash\forall[c<:B_1] E_1 <: \forall[c<:B_2] E_2}

\end{multicols}

\vspace{-1.5em}
\caption{Typing rules of System Capless.}\label{fig:all-typing}

\end{figure*}

Figure~\ref{fig:all-typing} presents the typing rules of System Capless.
The typing judgement has the form $C;\Gamma\vdash t : E$,
where $C$ is the \emph{use-set} of the term,
$\Gamma$ is the typing context,
$t$ is the term being typed,
and $E$ is the derived type.
\emph{Use-set} over-approximates the set of capabilities
that this term may use during its evaluation to an answer.

\subsection{Evaluation Rules}
\label{sec:appendix-reduction}

\begin{figure*}[htbp]
\scriptsize

\flushleft{\textbf{Syntax}}

\begin{align*}
  \Sigma\coloneqq\ &           \tag*{\textbf{Store}}\\
  &\emptyset                    \tag*{empty}\\
  &\Sigma, \textbf{\textsf{val}}\, x\mapsto v                    \tag*{val binding}\\
  \Psi\coloneqq\ &           \tag*{\textbf{Evaluation Context}}\\
  &[]                    \tag*{hole}\\
  &\textsf{let}\ x = \Psi \ \textsf{in}\ t           \tag*{let}\\
  &\textsf{let}\ \langle c,x \rangle = \Psi \ \textsf{in}\ t           \tag*{ex. let}\\
\end{align*}

\flushleft{\textbf{Reduction \quad $\langle\Sigma\,\vert\, t\rangle\longrightarrow\langle\Sigma'\,\vert\, t'\rangle$}}

\infrule[\ruledef{apply}]
{\Sigma(x) = \lambda(z: T)t}
{\langle\Sigma\,\vert\,\Psi[ x\,y ]\rangle \longrightarrow \langle\Sigma\,\vert\,\Psi[ [z:=y]t ]\rangle}

\infrule[\ruledef{tapply}]
{\Sigma(x) = \lambda[X<:S]t}
{\langle\Sigma\,\vert\,\Psi[ x[S'] ]\rangle \longrightarrow \langle\Sigma\,\vert\,\Psi[ [X:=S']t ]\rangle}

\infrule[\ruledef{capply}]
{\Sigma(x) = \lambda[c<:B]t}
{\langle\Sigma\,\vert\,\Psi[ x[C] ]\rangle \longrightarrow \langle\Sigma\,\vert\,\Psi[ [c:=C]t ]\rangle}

\infax[\ruledef{rename}]
{\langle\Sigma\,\vert\,\Psi[ \textsf{let}\ x = y \ \textsf{in}\ t ]\rangle\longrightarrow
 \langle\Sigma\,\vert\,\Psi[ [x:=y]t ]\rangle}

\infax[\ruledef{rename-e}]
{\langle\Sigma\,\vert\,\Psi[\textsf{let}\ \langle c, x \rangle =\langle C,y \rangle \ \textsf{in}\ t]\rangle\longrightarrow
 \langle\Sigma\,\vert\,\Psi[ [x:=y][c:=C]t]\rangle}

\infax[\ruledef{lift}]
{\langle\Sigma\,\vert\,\Psi[ \textsf{let}\ x = v\ \textsf{in}\ t ]\rangle \longrightarrow
 \langle\Sigma, \textbf{\textsf{val}}\, x\mapsto v\,\vert\,\Psi[ t ]\rangle}

\vspace{-1.5em}
\caption{Evaluation rules of System Capless.}\label{fig:reduction}

\end{figure*}

Figure~\ref{fig:reduction} presents the evaluation rules of System Capless.
The rules follow a small-step operational semantics with a store $\Sigma$ for value bindings
and an evaluation context $\Psi$ for specifying the evaluation order.
The key addition from System $\textsf{CC}_\textsf{formal}$ is the \ruleref{capply} rule for capture application
and the \ruleref{rename-e} rule for existential unpacking.
 
\end{document}